\documentclass[aps,prl,twocolumn,groupedaddress]{revtex4}
\usepackage{graphicx}
\usepackage{epstopdf}
\usepackage{bm}
\usepackage{amssymb}
\usepackage{psfrag}
\usepackage{color}
\usepackage{xcolor}
\usepackage{amsbsy}      
\usepackage{amsmath,amsthm}
\usepackage{mathtools}
\usepackage{dsfont}
\graphicspath{{./figures/}}
\usepackage{newfloat}
\DeclareFloatingEnvironment[name={Fig. D}]{suppfigure}

\usepackage{natbib} 
\def\dd{\mbox{d}}

\begin{document}

\title{Density- and elongation speed-dependent error correction 
in RNA polymerization}

\author{Xinzhe Zuo$^{1}$ and Tom Chou$^{1,2}$} \affiliation{$^{1}$Department of
  Mathematics, UCLA, Los Angeles, CA 90095-1555 \\$^{2}$Department of
  Computational Medicine, UCLA, Los Angeles, CA 90095-1766}




\begin{abstract}
Backtracking of RNA polymerase (RNAP) is an important pausing
mechanism during DNA transcription that is part of the error
correction process that enhances transcription fidelity. We model the
backtracking mechanism of RNA polymerase, which usually happens when
the polymerase tries to incorporate a mismatched nucleotide
triphosphate. Previous models have made simplifying assumptions such
as neglecting the trailing polymerase behind the backtracking polymerase
or assuming that the trailing polymerase is stationary.  We derive
exact analytic solutions of a stochastic model that includes locally
interacting RNAPs by explicitly showing how a trailing RNAP influences
the probability that an error is corrected or incorporated by the
leading backtracking RNAP. We also provide two related methods for
computing the mean times to error correction or incorporation given an
initial local RNAP configuration.
%
\end{abstract}
\maketitle

\section{Introduction}
Transcription is the first step of DNA-based gene expression. During
the process, an RNA polymerase (RNAP) enzyme binds and separates the
ds-DNA, forming a transcription bubble at a promoter site.  As the
RNAP and bubble move along the DNA, additional RNAPs can initiate new
bubbles at the empty promoter site.  Each RNAP processes along the DNA
up to the termination site, adding nucleotides to the 3' end of the
newly formed RNA transcript along the way.  The RNAP molecules and the
bubbles surrounding them form an exclusionary zone similar to that
seen in a chain of ribosomes translating mRNA during protein
production. Thus, it is natural to apply stochastic models such as the
totally asymmetric exclusion process (TASEP) originally developed for
studying mRNA translation
\cite{derrida1993exact,macdonald1968kinetics,derrida1993exact1D,lakatos2003totally,chou2004clustered,zia2011modeling,erdmann2018hydrodynamic}
to the DNA transcription process.

While DNA replication by DNA polymerase results in an error rate of
$10^{-8}$ to $10^{-10}$ per base pair
\cite{lynch2011lower,lang2008estimating,zhu2014precise}, RNA
polymerase has a much higher error rate of $10^{-5}$ to $10^{-6}$ per
base pair \cite{gout2013large,lynch2010evolution,shaw2002use}.
%
%
Since some RNAs are present at a level of less than one molecule per
cell in microbes \cite{pelechano2010complete} and in embryonic stem
cells \cite{islam2011characterization}, a gene may be represented by a
single mutated RNA transcript. Therefore, the fidelity of
transcription plays an important role in faithful gene expression.

RNAPs are sometimes interrupted by pauses
\cite{bai2006single,sydow2009rna}.  Krummel observed irregular DNA
footprints suggesting that RNAP shrinks and expands during the
elongation process \cite{krummel1992structural}. From this
observation, an ``inchworming" model for the elongation of RNAP was
developed \cite{nudler1995coupling}. However, later experiments
suggested that the inchworming phenomenon was actually the RNAP
complex traveling back and forth along the DNA template
\cite{nudler1997rna,komissarova1997rna,shaevitz2003backtracking}.  Now
known as RNAP backtracking, this important pausing mechanism aids
proofreading and fidelity of the transcription process.  Backtracking
strongly depends on the stability of the RNA/DNA hybrid in the
transcription bubble; the weaker the hybrid, the higher the
probability for backtracking \cite{nudler1997rna}.  Hence, when a
wrong nucleotide triphosphate (NTP) is added to the transcript, the 3'
end of the RNAP is frayed, which induces backtracking.

\begin{figure}[t]
\begin{center}
\includegraphics[width=3.4in]{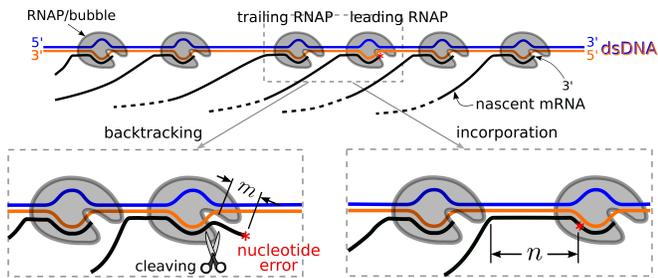}
\end{center}
\caption{Schematic of interacting RNAPs. A chain of transcribing RNAPs
  processing along DNA. Each RNAP forms a local DNA bubble through
  which one strand is copied to RNA. If an erroneous nucleotide (red
  asterisk) is recruited, the RNAP can either incorporate the error
  (lower right inset) or backtrack along the DNA (lower left
  inset). During backtracking, the RNAP loosens its grip, the active
  site reverses along the transcribed RNA in a random-walk-like
  fashion, and can cleave off the misincorporated nucleotide (lower
  left inset). In the backtracking state, cleaving near the 3' end of
  the mRNA can excise the error. The distance of the backtracking RNAP
  behind the erroneous 3' end of the mRNA is denoted $m$, while the
  genomic distance between the erroneous 3' end to the trailing
  polymerase is denoted $n$.}
\label{FIG1}
\end{figure}

During backtracking, the 3' end of the RNA disengages with the RNAP
catalytic site, rendering the RNAP complex inactive but stable
\cite{nudler1997rna,komissarova1997transcriptional}. Fig.~\ref{FIG1}
depicts a chain of RNAPs, their associated nascent RNA transcripts,
and one erroneous nucleotide (red asterisk). We assume that once a
wrong nucleotide is added to the catalytic site, the RNAP enters a
backtracking state during which it can moves backwards relative to
both the DNA and the RNA transcript without depolymerizing the
transcript. As a result, the 3' end of the RNA transcript now extrudes
out of the RNAP.  There are two competing processes for the RNAP to
exit the backtracking state, as depicted in the lower insets of Fig.~\ref{FIG1}.
%
%
In one, the RNAP can perform a random walk on the DNA template until
realignment occurs
\cite{galburt2007backtracking,depken2009origin,hodges2009nucleosomal}
and the erroneous nucleotide is incorporated into the transcript. In
the other, a segment of transcript associated with backtracking RNAP
can be cleaved so that a new RNA 3' end which aligns with the active
site is created
\cite{kuhn2007functional,chedin1998rna,orlova1995intrinsic}.  In
eukaryotic and prokaryotic cells, transcript cleavages are enhanced by
cleavage factors TFIIS \cite{reinberg1987factors,izban1992rna} and
GreA/GreB \cite{borukhov1993transcript}, respectively. Cleavage of the
mismatched nucleotide before incorporation allows the transcript under
construction to be corrected \cite{sydow2009rna}.

Previous theoretical studies have studied in detail the backtracking
kinetics of a single RNAP as elongation occurs, giving rise to
non-Poissonian pause times and bursty mRNA production
\cite{LIVERPOOL2008}. Edgar et al. \cite{roldan2016stochastic} also
examined the mean depth and time of the backtracking in both discrete
and continuous cases in semi-infinite chain, while Sahoo and Klump
\cite{sahooklump2013backtracking} studied the accuracy of the
transcription in the context of a single RNAP. Both studies assumed
that the trailing RNAP is stationary. However, when the leading RNAP is in
a backtracking state, the trailing RNAP is not stationary and would
most likely be in the active processing state closing the gap and/or
``pushing'' the leading RNAP forward
\cite{epshtein2003cooperation,epshtein2003transcription,jin2010synergistic,nudler2012rna}.
In this paper, we derive and solve a discrete stochastic model that
incorporates a trailing RNAP that closes in on the leading one. This
allows us to understand how interactions between neighboring RNAPs
influence the probabilities and timescales of error correction.  Our
approach also provides a starting point for collective, many-body
models of transcription.

\section{Local stochastic model}

\begin{figure}[t]
\begin{center}
\includegraphics[width=2.7in]{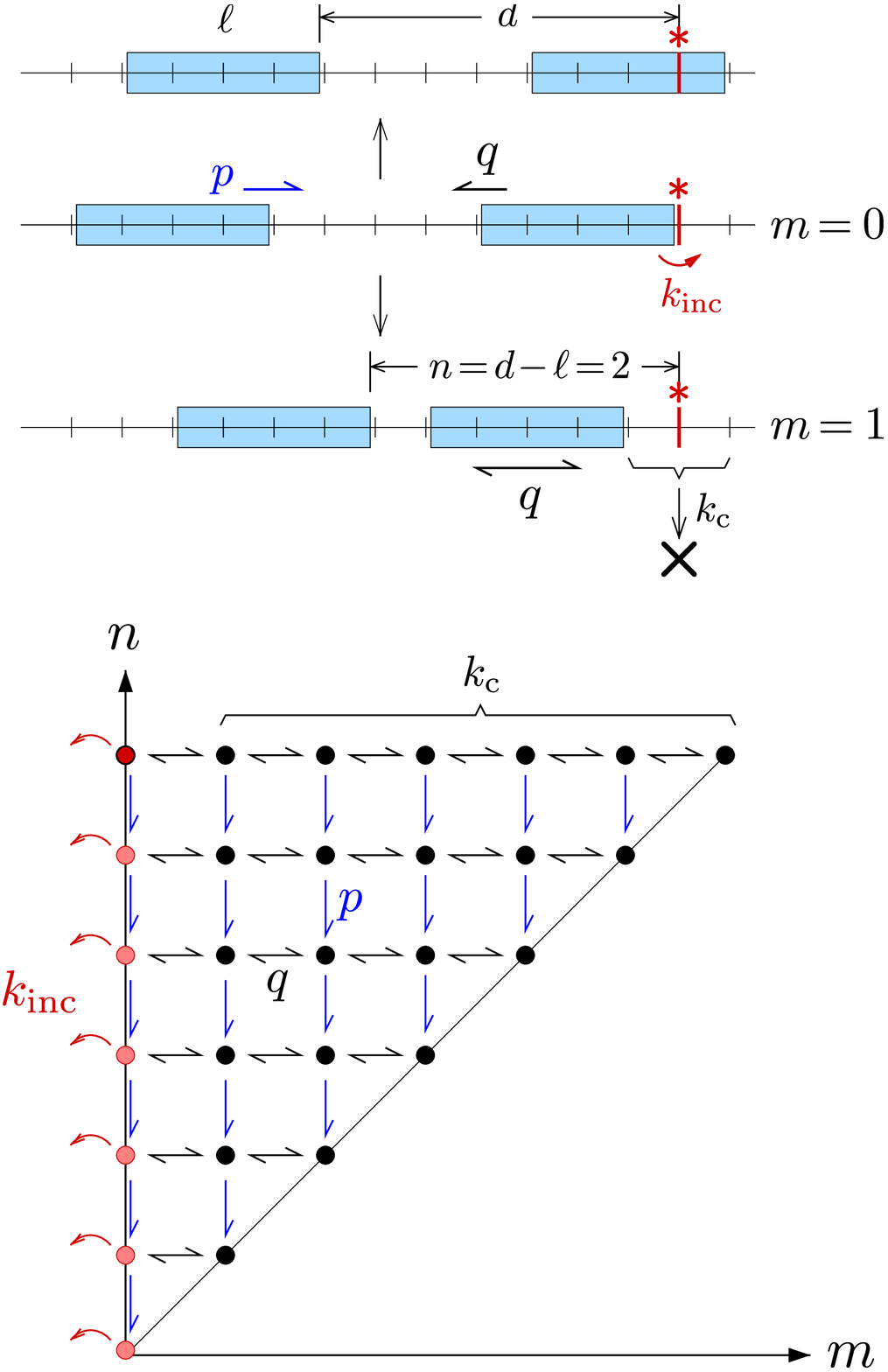}
\end{center}
\caption{Top panel: Local stochastic model for RNAP
  backtracking. Bottom panel: State-space of RNAP immediately
  following a wrong nucleotide addition. When this happens, assume
  that the open distance to the second RNAP is $n(t=0)\equiv N$. While
  in a backtracking state, the leading RNAP diffusively hops forward
  and backward with rate $q$. The trailing RNAP hops forward with rate
  $p$.  There are two mechanisms to escape from the backtracking
  state: incorporation of the wrong nucleotide with rate $k_{\rm inc}$
  while in states $(m=0,n)$, and cleavage with rate $k_{\rm c}$ while
  in states $(1 \leq m \leq n)$. In the diagram, the states indicated
  by black dots can undergo cleavage while those indicated by red dots
  can incorporate the error.}
\label{FIG2}
\end{figure}

Consider RNAPs with effective size $\ell$ (which included the
associated transcription bubble) that normally process along the gene
at rate $p$ as shown in Fig.~\ref{FIG2}. For clarity, the RNA
transcripts emanating from the RNAPs are not shown.  We now focus on
two adjacent RNAPs: a leading one that has just recruited a wrong
nucleotide (at the position marked by the red asterisk) and a trailing
one just downstream of the leading RNAP.  The nucleotide mismatch
promotes transition of the leading RNAP into the backtracking state
\cite{sydow2009rna,toulokhonov2007central}.  In this state, the
leading RNAP will have a smaller rate $k_{\rm inc}$ of moving forward
and incorporating the erroneous NTP or can undergo a symmetric random
walk with rate $q$ in the space between the trailing RNAP and the
realignment position (red asterisk). During the diffusive motion, the
end fragment of the transcript can also be cleaved with rate $k_{\rm
  c}$, removing the erroneous NTP and rescuing the leading RNAP from
the backtracking state as it resumes elongation.  At the same time,
the trailing RNAP is still moving forward with rate $p$ if it is
unblocked by the leading RNAP.

Define $m$ to be the distance between the leading RNAP and the
realignment position. Let $n$ be the sum of the distances that each
RNAP can move, i.e. it is the sum of the distance between the trailing
RNAP and the leading RNAP and $m$. If we set $d$ to be the distance
between the trailing RNAP and the realignment position, then by
definition we have $n=d-\ell$. Start the system of two RNAPs when the
leading one has just added a wrong nucleotide but has not incorporated
it yet. The evolution of the system can be described by the state
diagram in Fig.~\ref{FIG2}.

For the interior points, $m \geq 1$ and $n > m$, 

\begin{align}
{\dd P_{n}(m,t) \over \dd t} = & -(k_{\rm c}+2q+p)P_{n}(m,t) \nonumber \\
\: & +pP_{n+1}(m,t) +qP_{n}(m+1,t) \nonumber \\
\: & + qP_{n}(m-1,t).
\label{ME_INTERIOR}
\end{align}
For the boundary states $m=0$, $n\geq 1$, 
\begin{align}
{\dd P_{n}(0,t)\over \dd t} = & -(k_{\rm inc}+p+q)P_{n}(0,t)+qP_{n}(1,t) \nonumber \\
\: & + pP_{n+1}(0,t),
\label{ME_0}
\end{align}
while the probabilities of the edge states $m=n$ obey 

\begin{align}
{\dd P_{n}(n,t)\over \dd t} = & pP_{n+1}(n,t) + qP_{n}(n-1,t) \nonumber \\
\: &  - (k_{\rm c}+q)P_{n}(n,t),
\label{ME_N}
\end{align}
and that of the corner point obeys

\begin{equation}
{\dd P_{0}(0,t)\over \dd t} = pP_{1}(0,t) - k_{\rm inc}P_{0}(0,t).
\label{ME_00}
\end{equation}

The initial condition, defined at the instant a wrong nucleotide is
added is $P_{n}(m, t=0) = \mathds{1}(m,0)\mathds{1}(n,N)$.  Solution
of Eqs.~\ref{ME_INTERIOR}, \ref{ME_0}, and \ref{ME_N} yields the
probability the system is in state ($m, n$) at time $t$.

\subsection{Iterative Solution for $n=N$}

First, consider an initial fixed distance $n=N$ between the trailing
RNAP and the site of misincorporation (see top panel,
Fig.~\ref{FIG2}). Since the forward motion of the second RNAP is
unidirectional, the $n=N$ chain provides a source of probability flux
into the $n=N-1$ chain.  From the probabilities distributed across the
$n=N$ chain, we can calculate the time-dependent probability fluxes
that drive the dynamics of the $n=N-1$ chain, and so on.

By defining the Laplace transform $\tilde{P}_{n}(m,s)
=\int_{0}^{\infty}e^{-st}P_{n}(m,t)\dd t$ and taking the Laplace
transform of Eq.~\ref{ME_0}, we first find $\tilde{P}_{N}(1,s)$ in
terms of $\tilde{P}_{N}(0,s)$ and successively substitute into
Eq.~\ref{ME_INTERIOR} to find for $0 \leq m \leq N$

\begin{equation}
\tilde{P}_{N}(m,s) = {D_{m-1}\over q^{m}}\tilde{P}_{N}(0,s)
-{D_{m-1}\over q^{m}}\sum_{k=0}^{m-1}{q^{2k} \over 
D_{k-1}D_{k}},
\label{PNP0}
\end{equation}
where the coefficients $D_{m}$ obey

\begin{equation}
D_{m+1} = (s+ k_{\rm c}+2q+p)D_{m}-q^{2}D_{m-1}, \quad m\geq 0.
\label{DRECURSION}
\end{equation}
To determine $\tilde{P}_{N}(0,s)$ and close the system, we apply the
boundary condition at the end of the chain (Eq.~\ref{ME_N}) to find
$\tilde{P}_{N}(N,s) = q\tilde{P}_{N}(N-1)/( s+k_{\rm c} + q)$.
%
%
Upon using Eq.~\ref{PNP0} for $\tilde{P}_{N}(N,s)$ and
$\tilde{P}_{N}(N-1,s)$, we find

\begin{align}
& D_{N-1}\left[\tilde{P}_{N}(0,s) - 
\sum_{k=0}^{N-1}{q^2k \over D_{k-1}D_{k}}\right] \nonumber \\
& \hspace{1.2cm} = {q^{2}D_{N-2}\over s+k_{\rm c}+q}\left[\tilde{P}_{N}(0,s)
-\sum_{k=0}^{N-2}{q^{2k} \over D_{k-1}D_{k}}\right],
\end{align}
from which we find $\tilde{P}_{N}(0,s)$ explicitly

\begin{align}
\tilde{P}_{N}(0,s) = & \sum_{k=0}^{N-2}{q^{2k} \over
  D_{k-1}D_{k}} \nonumber  \\ 
\: & + {1\over
  D_{N-2}}\left[{q^{2(N-1)}(s+k_{\rm c}+q)\over (s+k_{\rm
      c}+q)D_{N-1}-q^{2}D_{N-2}}\right].
\label{P0SOLN}
\end{align}

The recursion in $D_{m}$ starts with $D_{-1}\equiv 1$, $D_{0}=s+k_{\rm
  inc}+p+q$.  To find an explicit expression for $D_{m}$
we use the generating function $G(z) \equiv \sum_{m=0}^{\infty}D_{m}z^{m}$ to
convert Eq.~\ref{DRECURSION} to

\begin{equation}
{1\over z^{2}}\left[G(z) - D_{0}-D_{1}z\right] = {A \over
  z}\left[G(z)-D_{0}\right]-q^{2}G(z),
\end{equation}
which is solved by

\begin{align}
G(z) & = {D_{0}+zD_{1} -zAD_{0} \over q^{2}z^{2}-Az + 1} \nonumber \\
\: & = {D_{0}(1-zA) + zD_{1}\over q^{2} (z_{+}-z_{-})}\left(
{1\over z-z_{+}} - {1\over z-z_{-}}\right),
\end{align}
where $z_{\pm}>0$ and $z_{+}> z_{-}$:

\begin{equation}
z_{\pm} = {(s+\lambda_{\rm c})\over 2q^{2}}\left[1\pm \sqrt{1-
{4q^{2}\over (s+\lambda_{\rm c})^{2}}}\right].
\end{equation}
By using $(1-z/z_{\pm})^{-1} = \sum_{k=0}^{\infty}(z/z_{\pm})^{k}$, 
%
%
we find the power series of $G(z)$ about $z=0$ 
(or use the inverse \texttt{Z}-transform) to 
find

\begin{widetext}
\begin{equation}
G(z) = {D_{0}-(D_{1}-AD_{0})z\over (z_{+}-z_{-})}\left[{1\over z_{-}}
\sum_{m=0}^{\infty}\left({z\over z_{-}}\right)^{m} - 
 {1\over z_{+}}\sum_{m=0}^{\infty}\left({z\over z_{+}}\right)^{m}\right],
\end{equation}
and hence an explicit expression for $D_{m}$:

\begin{equation}
D_{m\geq 2} = {D_{0}\over q^{2}(z_{+}-z_{-})}\left({1\over
  z_{-}^{m+1}}-{1\over z_{+}^{m+1}}\right)+ {(D_{1}-AD_{0})\over
  q^{2}(z_{+}-z_{-})}\left({1\over z_{-}^{m}}-{1\over z_{+}^{m}}\right).
\end{equation}
\end{widetext}
We can substitute $\tilde{P}_{N}(0,s)$ from Eq.~\ref{P0SOLN} into
Eq.~\ref{PNP0} and use the above expression for $D_{m}$ to find an
explicit solution to $\tilde{P}_{N}(m,s)$. The above results assume a
fixed trailing RNAP but will be used to construct the full solution in
the presence of a forward-moving trailing RNAP. Nonetheless, this
one-row ($n = N$) approximation provides a lower bound on the
probability that the wrong nucleotide is incorporated.

\subsection{Closing trailing particle}

Since elongation is irreversible, the system is feed-forward; that is,
the probabilities in the $n=N$ layer feed into the $n=N-1$ layer, and
so on.  The probability flux from the $n$ chain into each state $m
\leq n-1$ of the $n-1$ chain is $\tilde{J}_{n-1}(m,s)=
p\tilde{P}_{n}(m,s)$.  Thus, the probabilities within the $n-1$ chain
can be described by a recursion relation with an additional source of
probability from the $n$ layer:

\begin{align}
q\tilde{P}_{n}(m+1,s) = & {D_{m}\over D_{m-1}}\tilde{P}_{n}(m,s) \nonumber  \\
\: & - {q^{m}\over D_{m-1}}\sum_{k=0}^{m}D_{k-1} q^{-k}
\tilde{J}_{n}(k,s),
\label{Pn}
\end{align}
where $\tilde{J}_{n}(k,s) = p\tilde{P}_{n+1}(k,s)$.  Equation~\ref{Pn}
can be easily recursed to find an explicit expression for
$\tilde{P}_{n}(m,s)$ in the $n$ layer:

\begin{equation}
\tilde{P}_{n}(m,s)={D_{m-1}\over q^{m}}\left[\tilde{P}_{n}(0,s) - 
\sum_{\ell=0}^{m-1} q^{\ell} \tilde{Q}_{n}(\ell,s)\right],
\label{LAYER}
\end{equation}
where

\begin{align}
\tilde{Q}_{n}(\ell,s) & = {q^{\ell} \over
  D_{\ell}D_{\ell-1}}\sum_{k=0}^{\ell}D_{k-1} q^{-k}\tilde{J}_{n}(k,s).
%
\end{align}
We now Laplace-transform boundary condition in Eq.~\ref{ME_N}
to find 

\begin{equation}
\tilde{P}_{n}(n,s) = {\tilde{J}_{n}(n,s) +q\tilde{P}_{n}(n-1,s) 
\over s+q+k_{\rm c}}.
\label{eq:P_n_n_s}
\end{equation}
After using Eq.~\ref{LAYER} for $\tilde{P}_{n}(n,s)$ and 
$\tilde{P}_{n}(n-1,s)$ in Eq.~\ref{eq:P_n_n_s}, we can explicitly solve for 

\begin{widetext}

\begin{equation}
\displaystyle \tilde{P}_{n}(0,s) = {q^{n}\tilde{J}_{n}(n,s) + 
(s+q+k_{\rm c})D_{n-1}\sum_{\ell=0}^{n-1}q^{\ell}\tilde{Q}_{n}(\ell,s) 
- q^{2}D_{n-2}\sum_{\ell=0}^{n-2}q^{\ell}\tilde{Q}_{n}(\ell,s)
  \over (s+q+k_{\rm c})D_{n-1}-q^{2}D_{n-2}},
\label{eq:P_n_0_s}
\end{equation}
\end{widetext}
which we can use in Eq.~\ref{LAYER} to find an explicit expression for
$\tilde{P}_{n}(m,s)$.  Note that $\tilde{P}_{n}(m,s)$ depends on
$\tilde{Q}_{n}(\ell,s) \propto \tilde{J}_{n} = p \tilde{P}_{n+1}$, the
probabilities in the layer immediately above it.

\subsection{Outcome probabilities and times}

With the Laplace-transformed probabilities derived, we can calculate
the probabilities that the erroneous NTP is incorporated or cleaved.
The probability that the RNAP incorporates the wrong nucleotide by
time $t$ can be calculated by time-integrating the probability flux
\begin{align}
P_{\rm inc}(t) & =
k_{\rm inc}\sum_{n=0}^{N}\int_{0}^{t}\!\!\! P_{n}(m=0,t')\dd t'
\label{P_INC}
\end{align}
The final probability of wrong nucleotide incorporation
is $P_{\rm inc}(\infty) =  k_{\rm inc}\sum_{n=0}^{N}\tilde{P}_{n}(m=0,s=0)$,
while the total probability of cleaving is  $P_{\rm c}(\infty) = 1-P_{\rm inc}(\infty)$.

We can also define the density of incorporation times, conditioned on
incorporation of a wrong nucleotide, as $w(t) = k_{\rm
  inc}P_{n}(m=0,t)/P_{\rm inc}(\infty)$ and find the moments of the
conditioned incorporation time \cite{FPT}
\begin{align}
\mathds{E}[T_{\rm inc}^{\alpha}] & = {(-1)^{\alpha}k_{\rm inc}\over
  P_{\rm inc}(\infty)} \left[{\partial^{\alpha} \over \partial
    s^{\alpha}}\sum_{n=0}^{N}\tilde{P}_{n}(m=0,s)\right]_{s=0}\!\!\!. 
\label{Tinc}
%
\end{align}
%
%
%
Similarly, the moments of the times to cleavage (and correction of the
misincorporated nucleotide), conditioned on cleavage 
is

\begin{align}
\mathds{E}[T_{\rm c}^{\alpha}] & = {(-1)^{\alpha}k_{\rm c}\over 
P_{\rm c}(\infty)}
\left[{\partial^{\alpha} \over 
\partial s^{\alpha}}\sum_{n=0}^{N}\sum_{m=1}^{n}\tilde{P}_{n}(m,s)\right]_{s=0}\!\!\!.
\label{Tc}
%
\end{align}
Finally, the \textit{unconditional} resolution time, the time for the system to
either cleave or incorporate obeys

\begin{align}
\mathds{E}[T^{\alpha}] = & 
(-1)^{\alpha} k_{\rm inc}\left[{\partial^{\alpha} \over 
\partial s^{\alpha}}\sum_{n=0}^{N}\tilde{P}_{n}(m=0,s)\right]_{s=0} \nonumber \\
\: & \hspace{3mm}+ 
(-1)^{\alpha}k_{\rm c}\left[{\partial^{\alpha} \over 
\partial s^{\alpha}}\sum_{n=0}^{N}\sum_{m=1}^{n}\tilde{P}_{n}(m,s)\right]_{s=0}\!\!\!.
\label{T}
\end{align}
%

\section{Results and Discussion}

\begin{figure}[htb]
\begin{center}
\includegraphics[width=2.95in]{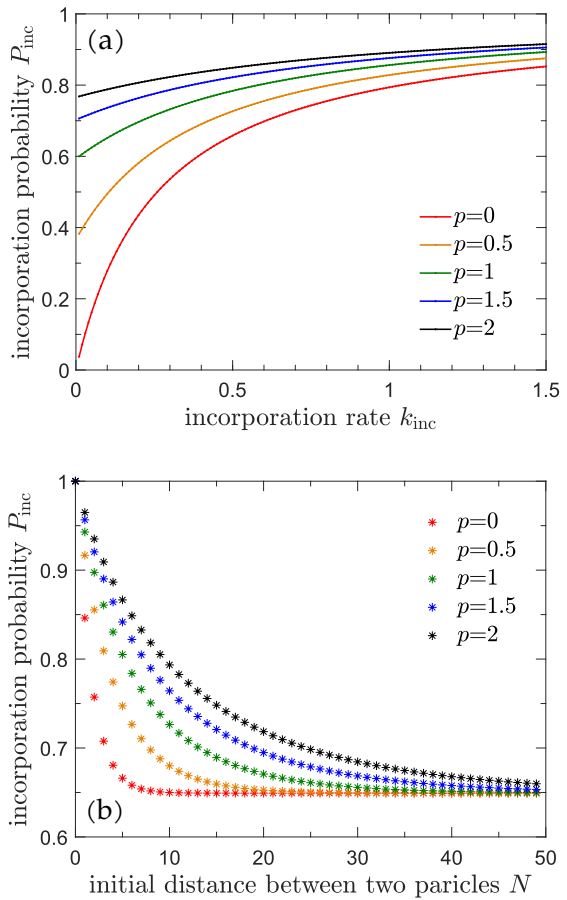}
\end{center}
    \caption{(a) Incorporation probability as a function of the error
      incorporation rate $k_{\rm inc}$. We measure all rates in terms
      of $q$, set $k_{\rm c}=0.1$ and explore different values of
      $p$. (b) Incorporation probability $P_{\rm inc}(\infty)$ as a
      function of initial separation $N$. As $N$ is reduced, the
      incorporation probability increases as the backtracking RNAP has
      less opportunity to cleave as it is more confined and spends
      more time abutted against the misincorporation site ($m=0$).
      For both plots, the initial RNAP separation $N=6$.}
    \label{FIG3}
\end{figure}

Henceforth, we will nondimensionalize time by $1/q$ and measure all
rates in terms of $q$. In Fig.~\ref{FIG3}(a-b), we use Eq.~\ref{P_INC}
to plot the final incorporation probability $P_{\rm inc}(\infty)$ as a
function of the incorporation rate $k_{\rm inc}$ and the initial RNAP
separation $N$ for different values of the trailing RNAP elongation
rate $p$.
Although $P_{\rm inc}(\infty)\propto k_{\rm inc}$, it increases
sublinearly with $k_{\rm inc}$ (Fig.~\ref{FIG3}(a)) because random
diffusion mitigates the incorporation by distributing the RNAP away
from the $m=0$ incorporation site. Nonetheless, as $k_{\rm inc}$
increases, the RNAP is more likely to incorporate the error.  For a
fixed $k_{\rm inc}$, having a faster elongation rate $p$ yields higher
incorporation probability since there is effectively less time for the
leading RNAP to cleave the erroneous nucleotide.


Fig.~\ref{FIG3}(b) shows that $P_{\rm inc}(\infty)$ converges to the
common value $P_{\rm inc}(\infty) \approx 0.65$ as $N\to \infty$. This
corresponds to an infinitely far trailing RNAP that will not influence
error correction of the leading RNAP. Note that $P_{\rm inc}(\infty)$
reaches the asymptotic value $0.65$ faster for smaller $p$. In all
cases, the final error incorporation probability increases with RNAP
translocation rate $p$ and can be thought of as a trailing RNAP
``pushing'' a backtracking-state (leading) RNAP to incorporate the
error.

\begin{figure}
    \begin{center}
    \includegraphics[width=3.2in]{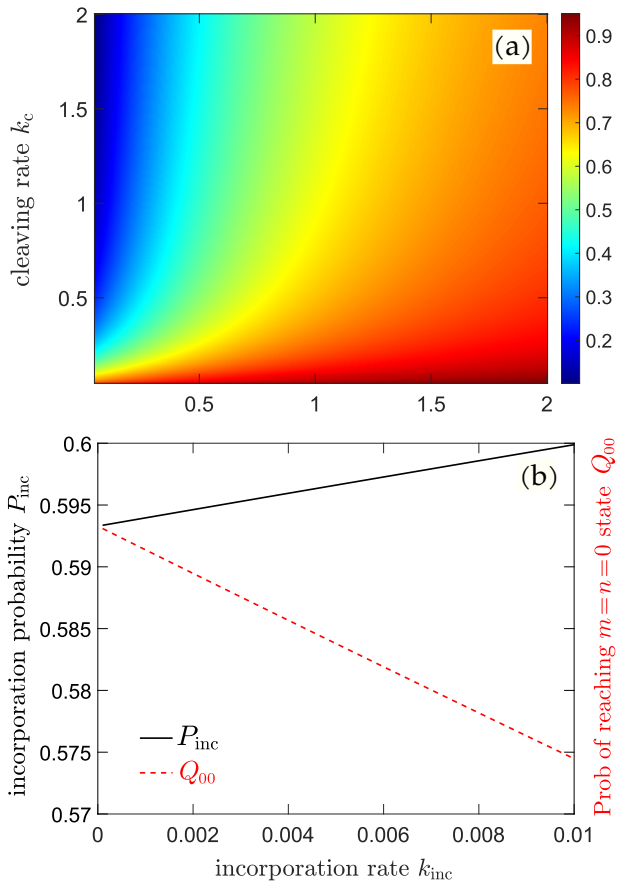}
\end{center}
    \caption{(a) Incorporation probability of the full model, plotted
      as a density with $p=1$ and an initial gap size $N=6$.  $P_{\rm
        inc}$ is monotonically decreasing (increasing) with increasing
      $k_{\rm c}$ ($k_{\rm inc}$). The iso-probability lines are
      approximately quadratic in $k_{\rm inc}$.  (b) For illustration,
      we keep $p=1$, set $k_{\rm c}=0.1$ and plot, in the small
      $k_{\rm inc}$ limit, the probability of incorporation $P_{\rm
        inc}$ and the probability $Q_{00}$ of reaching the compressed
      $m=n=0$ state. Note that $P_{\rm inc}$ approaches a finite value
      as $k_{\rm inc} \to 0$, because the $m=0$ states constitute a
      ``kinetic trap'' if the trailing RNAP abuts against the leading
      RNAP.}
\label{FIG4}
\end{figure}

In Fig.~\ref{FIG4}(a), we fix $p=1$, set the initial gap size $N=6$,
and plot $P_{\rm inc}(\infty)$ as a function of the cleavage rate
$k_{\rm c}$ and the incorporation rate $k_{\rm inc}$.
In Figs.~\ref{FIG4}(b) we show that the limiting behavior of $P_{\rm
  inc}(\infty)\nrightarrow 0$ as $k_{\rm inc} \rightarrow 0$.  We
define $Q_{00}\equiv k_{\rm inc}\tilde{P}_{0}(m=0,s=0)=
p\tilde{P}_{1}(m=0,s=0)$ as the probability that the trailing RNAP
contacts the leading RNAP at the realignment position (the probability
that the $m=n=0$ ``compressed'' state is reached).  As also shown in
Fig.~\ref{FIG4}(b), $Q_{00}\nrightarrow 0$ as $k_{\rm inc} \rightarrow
0$. Since in the $m=n=0$ state, the only way to escape from the
backtracking state is through incorporation, $P_{\rm inc}(\infty) \geq
Q_{00}$ because incorporation may still occur outside of the $m=n=0$
state.  As $k_{\rm inc} \rightarrow 0$, we expect that incorporation
can occur only when cleavage becomes impossible, which is the case in
the $m=n=0$ state, where the only way to escape the backtracking state
is through incorporation. Therefore, as shown in Fig.~\ref{FIG4}(b),
$P_{\rm inc}\rightarrow Q_{00}$ as $k_{\rm inc} \rightarrow 0$. This
limiting probability decreases as $k_{\rm c}$ increases or $p$
decreases as the $m=n=0$ state becomes less likely.

\begin{figure}[htb]
\begin{center}
   \includegraphics[width=3.1in]{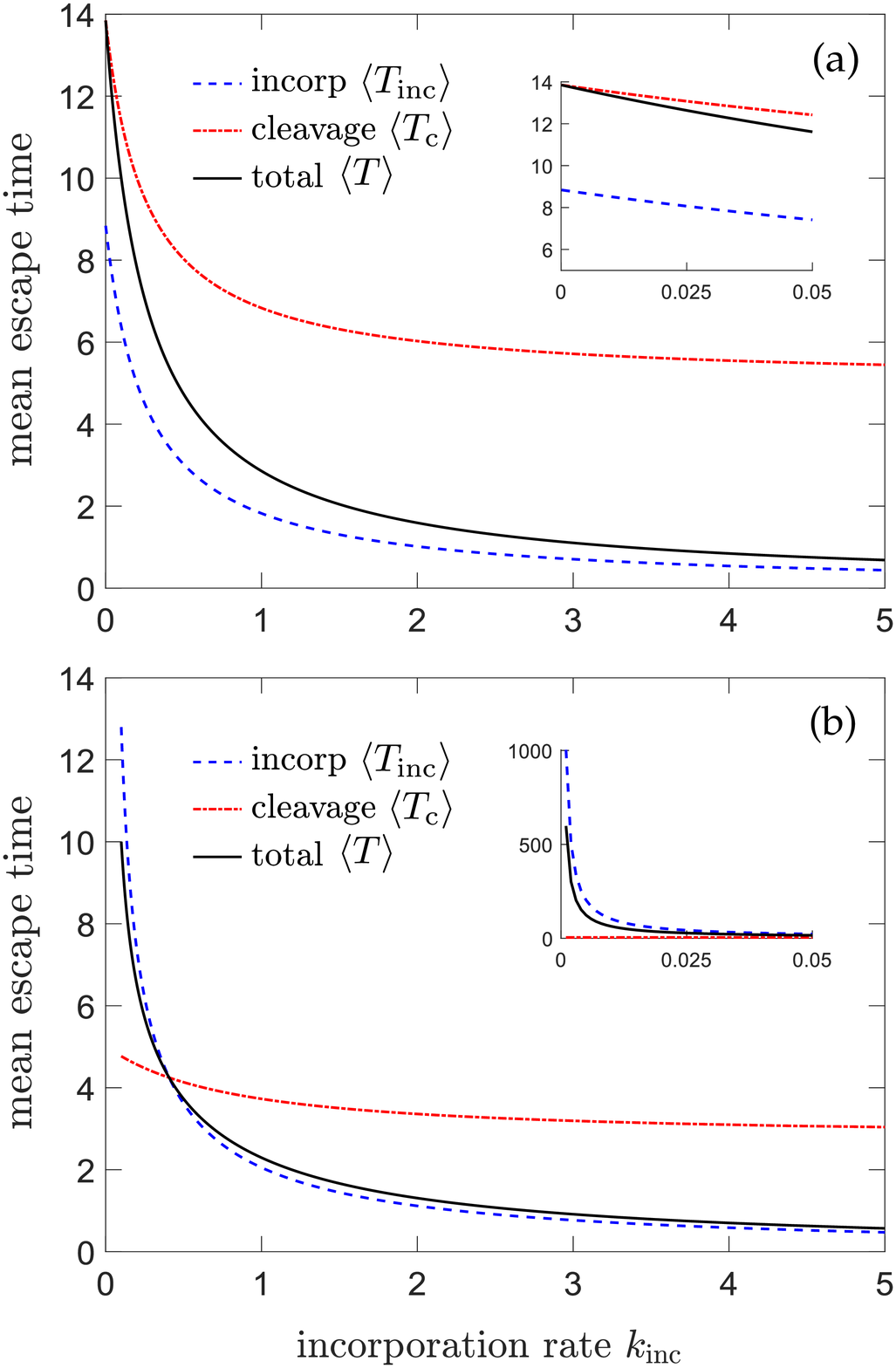}
\end{center}
    \caption{Mean times of incorporation ($\langle T_{\rm inc}\rangle$
      blue dashed), mRNA cleavage ($\langle T_{\rm c}\rangle$ solid
      red), and unconditional ( $\langle T\rangle$ solid black).  (a)
      Mean exit times in a system in which the trailing RNAP is
      stationary ($p=0$) with fixed $n=N=6$. (b) Mean exit times when
      the trailing RNAP advances (the full problem with $p=1$) with a
      starting distance of $N=6$. For both scenarios, we used $k_{\rm
        c}=0.1$}
    \label{fig:comparison_time}
\end{figure}

In Figs.~\ref{fig:comparison_time}(a-b) we use Eqs.~\ref{Tinc},
\ref{Tc}, and \ref{T} to plot the mean backtracking-state escape times
(first passage times), conditioned on incorporation, cleavage, or
neither.  When the trailing RNAP is stationary
(Fig.~\ref{fig:comparison_time}(a)), the mean escape time conditioned
on cleaving is always greater than the mean escape time conditioned on
incorporation.  Since the probability of incorporation vanishes as
$k_{\rm inc} \rightarrow 0$, the unconditioned mean escape time
approaches the mean time to cleave in this limit.  In the inset, we
see that both the conditioned and unconditioned mean exit times remain
finite as $k_{\rm inc} \rightarrow 0$ because when the trailing RNAP
is fixed, the system can always escape by cleaving.

We find qualitatively different behavior of mean escape times for the
full model in which the trailing RNAP is allowed to advance.
Fig.~\ref{fig:comparison_time}(b) shows the conditioned and
unconditioned mean exit times for a trailing RNAP with elongation rate
$p=1$. Here, the mean cleavage time is smaller than the mean
incorporation time if $k_{\rm inc}$ is sufficiently small. For $k_{\rm
  inc}\rightarrow 0$, as shown in the inset, both the unconditional
mean exit time and the mean incorporation time diverges. This
divergence arises since occupation of the $m=n=0$ state become more
likely and the mean incorporation time from this state scales as
$1/k_{\rm inc}$.
%
%
%
As the incorporation rate $k_{\rm inc}$ increases, the unconditioned
mean exit time approaches the mean incorporation time, which decreases
since it becomes increasingly likely for the leading particle to
incorporate the erroneous nucleotide.

In principle, all moments of exit times can be directly computed from
the $s$-dependence of $\tilde{P}_{n}(m,s)$ and Eqs.~\ref{Tinc},
\ref{Tc}, and \ref{T}. Here, we will simplify matters and only consider the
coefficient of variation (CV) of the exit times

\begin{equation}
{\rm CV} = \frac{\sqrt{\langle (T-\langle
    T\rangle)^{2}\rangle}}{\langle T\rangle}.
\label{CV}
\end{equation}
These CVs involve only the first and second moments of the escape
times and represent simple metrics that measures their deviation from
those of Poisson processes for which ${\rm CV}=1$.  Where appropriate,
we substitute $\langle T_{\rm inc}\rangle$ or $\langle T_{\rm
  c}\rangle$ for $\langle T\rangle$ above.
\begin{figure}[htb]
     \centering
         \includegraphics[width=3.2in]{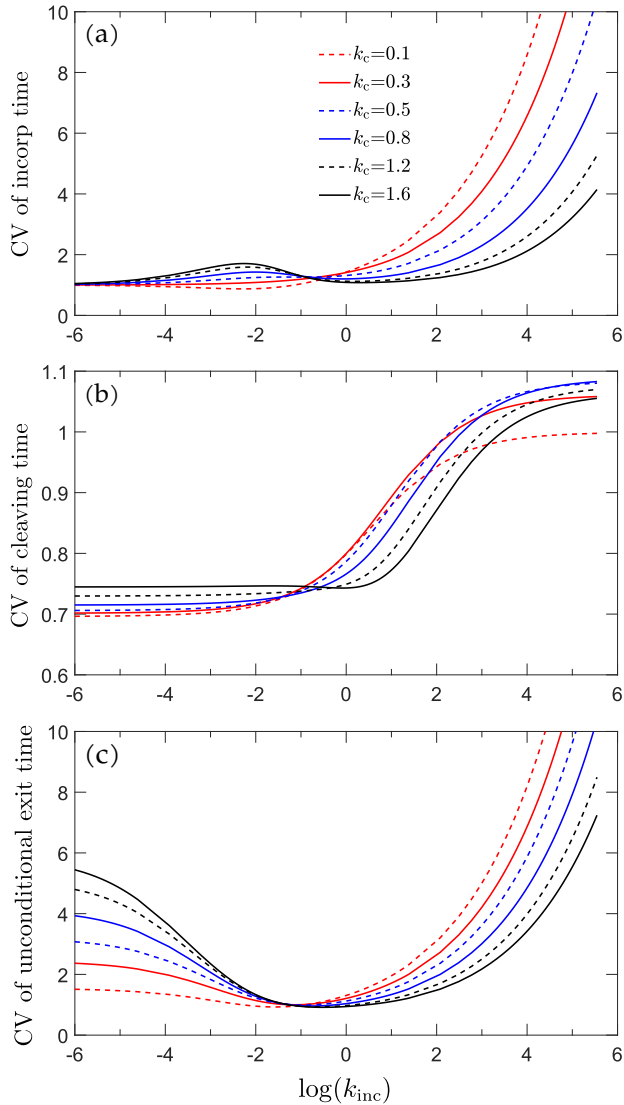}
        \caption{(a) The incorporation-time CV plotted as a function
          of $k_{\rm inc}$ for $p=1$, $N=6$, and various values of
          $k_{\rm c}$. (b) The CV of cleavage times. (c) General
          (unconditioned) escape times. These results show that the
          backtracking state (or ``pauses'') can exhibit
          non-Poissonian waiting times.}
    \label{FIG_CV}
\end{figure}
%
%

The escape-time CVs are plotted as functions of ${\rm log}(k_{\rm
  inc})$ in Fig.~\ref{FIG_CV}(a-c) for $N=6$, $p=1$, and various
$k_{\rm c}$.  The CV of the incorporation times shown in
Fig.~\ref{FIG_CV}(a) indicates a Poisson process in the $k_{\rm inc}
\to 0$ limit as incorporation becomes a rare event.  After peaking at
an intermediate $k_{\rm inc}$, the incorporation-time CV diverges as
$\sqrt{k_{\rm inc}}$ in the $k_{\rm inc}\to \infty$ limit.
Fig.~\ref{FIG_CV}(b) shows a cleaving-time CV that is below one for
small $k_{\rm inc}$ illustrating that cleaving can occur from
multiple, connected states.  For large $k_{\rm inc}$ and fixed $k_{\rm
  c}$, the CV remains near one (see (iii) below).  The asymptotic
limits in (iv) below are not depicted in (b). Finally, the
unconditioned exit time CV shown in Fig.~\ref{FIG_CV}(c) indicates a
large CV for small $k_{\rm inc}$ that approaches the Poisson limit
before increasing again at large $k_{\rm inc}$. These results for the
different waiting times indicate non-Poissonian behavior in RNAP
pausing as was found by Voliotis \textit{et al.}  under a different
stochastic model \cite{LIVERPOOL2008}.

The limiting behaviors of these CVs can be more simply understood and
approximated by considering a toy model consisting of only two states:
(1) an effective boundary $m=0$ state that can immediately incorporate
the error (with rate $k_{\rm inc}$) and (2) an effective interior
$m>0$ state that allows cleavage at rate $k_{\rm c}$. By lumping these
two classes of states into two states and labeling their probabilities
as $P_0(t)$ and $P_1(t)$, respectively, we can explicitly find
$\tilde{P}_0(s) = (s+k_{\rm c}+q)/[(s+k_{\rm inc}+q)(s+k_{\rm
    c}+q)-q^2]$ and $\tilde{P}_{1}(s) = q/[(s+k_{\rm inc} +q)(s+k_{\rm
    c}+q)-q^2]$ where the diffusive hopping rate $q$ in this
simplified model is the inter-state transition rate and the initial
condition is $P_{0}(t=0)=1$. From this toy model, we find

\begin{enumerate}
    \item[(i)] The incorporation-time CV $\to 1$ as $k_{\rm c} \to \infty$;
    \item[(ii)] The incorporation-time CV diverges as $\sqrt{k_{\rm
        inc}}$ for $k_{\rm inc} \to \infty$ (as shown in
      Fig.~\ref{FIG_CV}(a));
    \item[(iii)] The cleavage-time CV $\to 1$ when either $k_{\rm c}$ or
      $k_{\rm inc} \to \infty$ and the other is large compared to $q$;
    \item[(iv)] The CV of the cleavage times $\sim
      \sqrt{\frac{1+(k_{\rm c}/k_{\rm inc})^2}{(1+k_{\rm c}/k_{\rm
            inc})^2}}$ when $k_{\rm c}, k_{\rm inc}\to \infty$ with
      $k_{\rm c}/k_{\rm inc}$ fixed. For example, if $k_{\rm c} =
      k_{\rm inc} \to \infty$, the cleavage-time CV $\sim 1/\sqrt{2}$.
    \item[(v)] The CV of the overall (unconditioned) exit time diverges as
      $\sqrt{2k_{\rm inc}}/k_{\rm c}$ when $k_{\rm inc}\to \infty$;
    \item[(vi)] The CV of the overall exit time $\sim 1$ as $k_{\rm c},
      k_{\rm inc} \to \infty$ with $k_{\rm c}/k_{\rm inc}$ fixed.
\end{enumerate}
The predictions from this toy model conform to limiting results of
the full model shown in Fig.~\ref{FIG_CV}. Thus, these limiting
behaviors are independent of finite RNAP spacing $N$. The CVs provide
insight into the statistics of the exit times of a backtracking state
and will be useful in developing multi-RNAP exclusion models that can
allow for successive and/or multiple backtracking RNAPs.

\section{Summary and Conclusions}

RNA polymerase backtracking is an important mechanism for
transcription fidelity
\cite{nudler1995coupling,nudler1997rna,nudler2012rna} as it is an
intermediate step before cleavage of a misincorporated nucleotide.  To
study this process, we derived a stochastic model describing the
interactions between two processing RNAP enzymes after the leading one
has incorporated an erroneous nucleotide and transitioned into a
backtracking state. Previous studies have concluded that the trailing
RNAP will likely ``push'' the leading backtracking RNAP forward,
making it exit the backtracking state faster
\cite{nudler1995coupling,nudler1997rna,nudler2012rna}.  In our model,
we relax the assumption of a fixed-domain for the diffusing RNAP
particle, improving upon previous models
\cite{sahooklump2013backtracking}.  As the trailing RNAP moves
forward, the space available for the leading, backtracking RNAP to
move diminishes with time, allowing it to push the leading RNAP to
incorporate the error.

We used Laplace transforms to formally solve the three-parameter
stochastic model and found the probabilities for removing or
incorporating the erroneous nucleotide.  From analysis of our
solutions, we found that the ``pushed error incorporation'' effect
occurs only if the ratio of the incorporation rate $k_{\rm inc}$ to
the elongation rate $p$ is large enough. Otherwise, the system will
take a much longer time to exit the backtracking state (see
Fig.~\ref{fig:comparison_time}).  Our analyses also allowed for easy
computation of the conditioned mean times to error removal or
incorporation. Our main analytic approach also allows for the explicit
calculation of moments of removal and incorporation times.

Our model and the associated results provide the components needed in
more complete multi-RNAP descriptions. For example, a chain of RNAPs
may be described by an exclusion processes such as the TASEP, which
has been extensively used to describe mRNA \textit{translation}
\cite{GIBBS_1969,HYDRO_2006,zia2011modeling,erdmann2018hydrodynamic}. In such
many-body models, one could address multiple, simultaneously stalled
RNAPs and how their interactions affect probabilities of correction or
incorporation of each transcript.  A competition between transcription
fidelity and RNA production rate would be expected to arise and will
be the subject of future investigation.

\vspace{3mm}
\noindent \textbf{Acknowledgments:} This work was supported by grants
from the NIH through grant R01HL146552 (TC), the Army Research Office
through grant W911NF-18-1-0345 (TC), and the NSF through grant
DMS-1814364 (TC).


\bibliography{refs}

\section{Appendix: Alternate calculation of mean times}
Our method of solution requires solution of the recursion relations
for the probabilities as a function of all the rate parameters and the
Laplace-transformed time variable $s$. Thus, we explicitly carry all
time dependence throughout the calculation in terms of $s$. In the
end, we either set $s=0$ to find probabilities or take derivatives
with respect to $s$ and then set $s\to 0$ to find moments of the escape times.

However, if we are only interested in the mean condition times to
cleavage or incorporation, we can develop a simple coupled set of
recursion relations that can be easily evaluated numerically.

\subsection{Conditional mean times for a stationary trailing RNAP}
First, consider the case where the trailing RNAP is stationary -- we
treat the more general case of an advancing trailing particle in the
next subsection.  For an initial gap $N$ between the trailing RNAP and
the backtracking RNAP, we want to find the expected time $\langle
T_{\rm c}\rangle$ for the backtracking particle to cleave (correct)
the error, and the expected time $\langle T_{\rm c}\rangle$ to
incorporate the error given that cleavage or incorporation,
respectively, occurs.

A static trailing particle means that the system stays in
the first row of the state diagram in Fig.~\ref{FIG2} (bottom
panel). Let us label the states of the top row from left to right as
$B_0, B_1,\ldots, B_N$.  $B_0$ corresponds to the initial state where
the leading RNAP has just added a wrong nucleotide but has not yet
incorporated it. $B_N$ corresponds to the state where the leading RNAP
has backtracked a distance $N$ and abuts the trailing RNAP.

If the RNAP incorporates the error while in state $B_0$, then we
denote this state as $B_{-1}$; if the RNAP cleaves the error from
state $B_i$, $0<i\leq N$, then we denote this state as $B_{N+1}$.
Note that $B_{-1}$ and $B_{N+1}$ represent absorbing states associated
with error incorporation and error correction respectively.

Define $v_k = \mathbf{P}_k(X_T = B_{-1})$ as the probability that the
system reaches state $B_{-1}$ given that it started in state
$B_k$. These probabilities satisfy the recursion relations
\begin{align}
    v_1 &= (1+\gamma) v_0 - \gamma \nonumber \\
    v_{i} &= \frac{v_{i-1}}{\beta} - v_{i-2}, \,\,\, 2\leq i \leq N \nonumber \\
    v_{N} &= \frac{q}{q+k_{\rm c}} v_{N-1},
\label{eq:v}
\end{align}
where $\gamma = k_{\rm inc}/q$, $\beta = q/(2q+k_{\rm c})$,
$v_{-1}=1$, and $v_{N+1}=0$. One can show that the solution to
Eqs.~\ref{eq:v} is given by
\begin{align}
    v_i &= C_i v_0 - F_i \nonumber\\
    C_i &= x_1 \zeta_+^i + x_2 \zeta_-^i \nonumber \\
    F_i &= x_3 \zeta_+^i + x_4 \zeta_-^i,
\end{align}
where $\zeta_{\pm} = \frac{1\pm\sqrt{1-4\beta^2}}{2\beta}$, $x_1 =
\frac{\gamma+1-\zeta_-}{\zeta_+-\zeta_-}$, $x_2 = 1-x_1$, $x_3 =
\frac{\gamma}{\zeta_+-\zeta_-}$, and $x_4 = -x_3$. $v_0$ can be solved
by plugging in the above expressions for $v_i$ into the last equation
in \eqref{eq:v} which gives

\begin{equation}
    v_0 = \frac{(q+k_{\rm c}) F_{N} - q F_{N-1}}{(q+k_{\rm c})C_{N}-q C_{N-1}}. 
\label{eq:v0}
\end{equation}
One can check that Eq.~\eqref{eq:v0} and $k_{\rm
  inc}\tilde{P}(0,N,s=0)$ (Eq.~\ref{Tinc}) yield the same result when
we set the elongation rate $p=0$.

Next, we can study the mean escape time conditioned on
incorporation. Recall that the conditional expectation of a random
variable $X$ given an event $H$, where $P(H)>0$, is given by

\begin{equation}
    \mathbb{E}[X|H] = \frac{\mathbb{E}[ X \cdot \mathbf{1}_H]}{P(H)}\,
\label{eq:conditional_expectation}.
\end{equation}
We see directly from Eq.~\eqref{eq:conditional_expectation} that it is
necessary to require $P(H)>0$ for our equation to be well-defined
(interested readers can read about the Borel-Kolmogorov paradox for
the case $P(H)=0$). Following this idea, we define $u_i =
\mathbb{E}_i[T\cdot \mathbf{1}_A]$, where $T$ is the time to reach one
of the absorption states (i.e. unconditioned escape time),
$\mathbf{1}_A$ is the indicator function for the event
$A=\{X_T=B_{-1}\}$. The quantity we would like to find is
$\mathbb{E}_0[T|A]=u_{0}/v_{0}$, which is the mean escape time
conditioned on incorporation when the leading RNAP adds a wrong
nucleotide. Since we have already solved for $v_0$, it suffices to
find $u_0$. At each state $B_j$, there are rates for additional
transitions depending on $j$. We can view these as competing Poisson
processes. Suppose the rates are given by $r_{j_1},...,r_{j_d}$ at
state $B_j$. Then the mean waiting time for the next move is
$(\sum_{i=1}^d r_{j_i})^{-1}$. And the probability of choosing the
move with rates $r_{j_l}$ is simply $r_{j_l}/(\sum_{i=1}^d r_{j_i})$.
Therefore, we obtain
\begin{align}
    \mathbb{E}_0\left[T\cdot\mathbf{1}_A\right] = & \frac{k_{\rm inc}}{q+k_{\rm inc}}
\mathbb{E}_{-1}\left[(T+\frac{1}{q+k_{\rm inc}})\cdot \mathbf{1}_A\right] \nonumber \\
    & + \frac{q}{q+k_{\rm inc}}\mathbb{E}_{1}\left[(T+\frac{1}{q+k_{\rm inc}})
\cdot \mathbf{1}_A\right]\nonumber \\
    \mathbb{E}_i\left[ T\cdot\mathbf{1}_A\right]  = &
    \frac{q}{2q+k_{\rm c}}\mathbb{E}_{i-1}\left[(T+\frac{1}{2q+k_{\rm c}})
\cdot \mathbf{1}_A\right] \nonumber \\
    & +\frac{q}{2q+k_{\rm c}}\mathbb{E}_{i+1}\left[(T+\frac{1}{2q+k_{\rm c}})
\cdot \mathbf{1}_A\right] \nonumber \\
    \mathbb{E}_{N}\left[T\cdot\mathbf{1}_A\right] = &
    \frac{q}{q+k_{\rm c}}\mathbb{E}_{N-1}\left[(T+\frac{1}{q+k_{\rm c}})\cdot 
\mathbf{1}_A\right].
\label{eq:conditional_time}
\end{align}
After some algebra, we find
\begin{align}
    u_1 &=(1+\gamma)u_{0}-\frac{v_0}{q} \nonumber\\
    u_{i+1} &= \frac{u_i}{\beta}-\frac{v_i}{q}-u_{i-1} \nonumber \\
    u_{N} & = \frac{q}{q+k_{\rm c}}u_{N-1} + \frac{v_{N}}{q+k_{\rm c}}\,.
\label{eq:u}
\end{align}
The above recursion relation can be solved analytically as
\begin{equation}
    u_i = H_i u_0-(G_iv_0-K_i),
\label{eq:u_general_expression}
\end{equation}
where
\begin{align}
    H_n &= C_n\nonumber\\
    G_n &= s_1\zeta_+^n+s_2\zeta_-^n+ns_3\zeta_+^n+ns_4\zeta_-^n \nonumber\\
    K_n &=j_1\zeta_+^n+j_2\zeta_-^n+nj_3\zeta_+^n+nj_4\zeta_-^n \, \nonumber
\end{align}
and the coefficients $s_i$ and $j_i$ are given by 
\begin{align}
    s_1 &= -\frac{\beta(-1+\beta(1+2\beta+\gamma))}{q(1-4\beta^2)^{3/2}}\nonumber \\
    s_2 &=-s_1 \nonumber \\
    s_3 &= \frac{\beta(-1+\sqrt{1-4\beta^2}+2\beta(1+\gamma))}{q(2-8\beta^2)}\nonumber \\
    s_4&=\frac{\beta(1+\sqrt{1-4\beta^2}-2\beta(1+\gamma))}{q(-2+8\beta^2)}\nonumber \\
    j_1 &= -\frac{\beta^2\gamma}{q(1-4\beta^2)^{3/2}}\nonumber\\
    j_2 &= -j_1 \nonumber \\
    j_3 &=j_4 = \frac{\beta^2\gamma}{q-4q\beta^2}\,.
\end{align}
To find $u_0$, one can substitute
Eq.~\eqref{eq:u_general_expression} into the last equation of
\eqref{eq:u} to obtain

\begin{align}
u_0 =  &
\frac{[C_{N}+q(G_{N}-G_{N-1})+k_{\rm c}G_{N}]v_{0}}{q(C_{N}-C_{N-1})+k_{\rm c}C_{N}} \nonumber \\
\: & \hspace{6mm} - \frac{[F_{N}+q(K_{N}-K_{N-1})+k_{\rm c}K_{N}]}{q(C_{N}-C_{N-1})+k_{\rm c}C_{N}}.
\end{align}

Similarly, we can find the mean escape time conditioned on
cleaving. One can define $\bar{v}_i = 1-v_i$ as the probability of
starting from state $B_i$ and eventually ending in state $B_{N+1}$
(since there are only two absorbing states, a particle has to arrive
at one of them). One can check that

\begin{align}
    \bar{v}_0 &= q\bar{v}_1+(1-q-k_{\rm inc})\bar{v}_0 \nonumber\\
    \bar{v}_i&=\frac{k_{\rm c}\beta}{q} +\beta(\bar{v}_{i+1}+\bar{v}_{i-1})\nonumber \\
    \bar{v}_{N}&= \frac{k_{\rm c}}{q+k_{\rm c}}+\frac{q\bar{v}_{N-1}}{q+k_{\rm c}}\,.
\end{align}
If we define $\bar{u}_i=\mathbb{E}_i[T\cdot \mathbf{1}_{A^c}]$, as the
expected stopping time for the event $A^c=\{X_T=B_{N+1}\}$, we can
show that $\bar{u}_i$ satisfies the same equations as $u_i$ does with
$v_i$ changed to $\bar{v}_i$. The new recursion relation for
$\bar{u}_i$ can be expressed as
\[
\bar{u}_i = \bar{H}_i\bar{u}_0-(\bar{G}_iv_0-\bar{K}_i)\,.
\]
As before, we have $\bar{H}_i=C_i$, $\bar{G}_i=-G_i$. $\bar{K}_i$
takes on a different form because the recursive relation for
$\bar{K}_i$ yields one more root,

\[
\bar{K}_n = \bar{j}_1\zeta_+^n+\bar{j}_2\zeta_-^n+n\bar{j}_3\zeta_+^n
+n\bar{j}_4\zeta_-^n +\bar{j}_5\,,
\]
where
\begin{align}
    \bar{j}_1&=-\frac{\beta\left[1+\sqrt{1-4\beta^2}-2
\beta(2\beta-\sqrt{1-4\beta^2}+\gamma)\right]}{2q(1-4\beta^2)^{3/2}}\nonumber \\
    \bar{j}_2&=-\frac{\beta\left[-1+\sqrt{1-4\beta^2}+
2\beta(2\beta+\sqrt{1-4\beta^2}+\gamma)\right]}{2q(1-4\beta^2)^{3/2}}\nonumber \\
    \bar{j}_3&=\bar{j}_4=-\frac{\beta^2\gamma}{q-4q\beta^2}\nonumber \\
    \bar{j}_5&=\frac{\beta}{q-2q\beta}\nonumber\,.
\end{align}
We then find $\bar{u}_0$  
\begin{align}
    \bar{u}_0 = &  \frac{1}{(q+k_{\rm c})C_{N}-q C_{N-1}}\left[1-C_{N}v_0+F_{N} \right. \\
    & \left. -q(\bar{G}_{N-1}v_0-\bar{K}_{N-1})
+(q+k_{\rm c})(\bar{G}_{N}v_0-\bar{K}_{N})\right]. \nonumber
\end{align}

\begin{figure}[t]
\begin{center}
\includegraphics[width=3.2in]{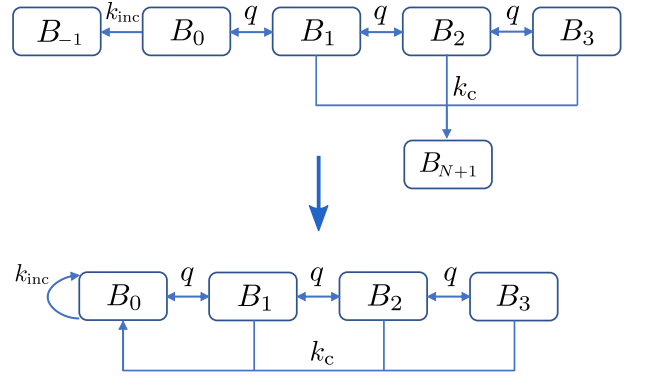}
\caption{Rewiring of states for computing mean exit times. The top
  panel shows the original transition of states. Recall that $B_{-1}$
  represents incorporation of error, and $B_{N+1}$ represents cleaving
  the error. In the bottom panel, all transitions to absorbing states
  ($B_{-1}$ and $B_{N+1}$) are rewired to the initial state $B_0$.}
\label{fig:rewire_state}
\end{center}
\end{figure}
With $u_0$ and $\bar{u}_0$ given, we are able to calculate the unconditioned
mean escape time $T_{\rm u}$ which is defined by

\begin{align}
    T_{\rm u} &= v_0 \mathbb{E}_i[T|{1}_A]+\bar{v}_0\mathbb{E}_i[T|{1}_{A^c}] \nonumber \\
    &= u_0 + \bar{u}_0\,. \label{eq:T_bt_unconditioned}
\end{align}
One can also apply the same arguments to $T_{\rm u}$ as we used for $u_i$
and $\bar{u}_i$ and seek $\mathbb{E}_i[T]$ instead of
$\mathbb{E}_i[T\cdot{1}_A]$.

Another way to calculate $T_{\rm u}$ was introduced by Hill
\cite{hill1988interrelations}. He showed that the unconditioned 
mean escape time can be calculated if we consider the steady state in a
transformed network without absorbing states. The transformed network
is obtained by rewiring the transitions to absorbing states to the
initial state in the original network.  For instance, the maximum
backtracking depth is set to be $N = 3$ in
Fig.~\ref{fig:rewire_state}, and all transitions to absorbing states
are rewired to the initial state

The probability distribution of the stationary state of the rewired
network can be found as $P_i=P_{N}L_{N-i}$, where $L_i=x'_1\zeta_+^i +
x'_2\zeta_-^i$, $x'_1 = (\lambda+1-\zeta_-)/(\zeta_+-\zeta_-)$,
$x'_2=1-x'_1$, $\lambda=k_{\rm c}/q$, and
\[P_{N} = \frac{1}{ x'_1 \frac{ 1-\zeta_+^{N+1}}{1-\zeta_+}
+x'_2\frac{1-\zeta_-^{N+1}}{1-\zeta_-}}\,.
\]
The unconditioned mean escape time is given by 
\begin{equation} 
T_{\rm u} = \frac{1}{P_0k_{\rm inc} + (1-P_0)k_{\rm c}}.
\label{eq:T_bt}
\end{equation}
One should note that we can get $T_{\rm u}$ for free by using
Eq.~\eqref{eq:T_bt_unconditioned} if we have the conditional
incorporation time $\mathbb{E}_i[T|{1}_A]$, the conditional cleavage
time $\mathbb{E}_i[T|{1}_{A^c}]$, the incorporation and cleavage
probability $v_0$, and $\bar{v}_0$. However, one cannot recover the
conditional mean times $\mathbb{E}_i[T|{1}_A]$ and
$\mathbb{E}_i[T|{1}_{A^c}]$ even if we know $T_{\rm u}$, $v_0$, and
$\bar{v}_0$ because essentially, we are trying to solve $x$ and $y$
from \cite{hill1988interrelations}
\[
xp+y(1-p)=c\,,
\]
which does not have a unique solution. 

\subsection{Mean conditional times for a trailing RNAP that advances}

To derive the incorporation probability when the trailing RNAP is
moving forward with elongation rate $p$, we use Fig.~\ref{FIG2} to
build our solution. Let $v(m,n)$ be the probability of incorporating
the error, given that the RNAP start at state $(m,n)$. Note that by
definition, $v(i,j)$ only makes sense for $0\leq i\leq j\leq N$, where
$N$ is the maximum backtracking depth, which is also the distance
between the trailing and leading RNAP when the backtracking dynamics
first started.

As a boundary condition, we have $v(0,0)=1$. This is because when the
leading RNAP is at the realignment position and there is no room
for backtracking, it can only incorporate the error and move
forward. Suppose now we have $v(i,j)$ for all $0\leq i \leq j$, we can
recursively build the solution for $v(i,j+1)$ for $0\leq i \leq
j+1$ via
\begin{align}
& v(0,j+1) = \frac{k_{\rm inc} + pv(0,j) + qv(1,j+1)}{k_{\rm inc} + p + q},\nonumber\\    
& v(i,j+1) = \frac{qv(i-1,j+1)+qv(i+1,j+1)+pv(i,j)}{k_{\rm c}+2q+p} \nonumber \\
& \hspace{6cm} \text{for all $1\leq i \leq j$} \nonumber \\
& v(j+1,j+1) =  \frac{qv(j,j+1)}{k_{\rm c}+q}. 
\label{eq:prob_moving_trailing}
\end{align}
We can use similar method as the previous section and study
(numerically) the mean escape times. Let $u(i,j)
=\mathbb{E}_{ij}[T\cdot \mathbf{1}_A]$, where $T$ is the time to reach
one of the absorption states, $\mathbf{1}_A$ is the indicator function
for the event $A=\{X_T=B_{-1}\}$ and the subscript $ij$ represents the
initial state $(m,n)$, with $0\leq j \leq N$ and $0 \leq i \leq
j$. The stochastic equations are given by
\begin{align}
u(0,0) = & \frac{1}{k_{\rm inc}},\nonumber \\
u(0,j) = & \frac{k_{\rm inc}}{(k_{\rm inc}+q+p)^2} +
\frac{pu(0,j-1) + qu(1,j)}{k_{\rm inc}+q+p} \nonumber \\
\: & \hspace{1cm} +\frac{pv(0,j-1)}{(k_{\rm inc}+q+p)^{2}} + \frac{qv(1,j)}{(k_{\rm inc}+q+p)^2},
\nonumber \\[12pt]
u(i,j) = & \frac{qu(i+1,j)}{2q+p+k_{\rm c}} + \frac{qv(i+1,j)}{(2q+k_{\rm c}+p)^{2}} \nonumber \\
\: & \hspace{1cm} + \frac{qu(i-1,j)}{2q+p+k_{\rm c}}+\frac{qv(i-1,j)}{(2q+k_{\rm c}+p)^{2}} \nonumber \\
\: &  \hspace{1cm} + \frac{pu(i,j-1)}{2q+p+k_{\rm c}} +\frac{pv(i,j-1)}{(2q+k_{\rm c}+p)^{2}}, 
\nonumber \\[12pt]
u(j,j) = & \frac{qu(j-1,j)}{q+k_{\rm c}} +\frac{qv(j-1,j)}{(q+k_{\rm c})^{2}}.
\label{eq:stoch_conditioned_time}
\end{align}
The derivation for $\tilde{u}(i,j) =\mathbb{E}_{ij}[T\cdot
  \mathbf{1}_{A^c}]$ is similar.  The linear system
\eqref{eq:prob_moving_trailing}-\eqref{eq:stoch_conditioned_time} can
be easily solved since the size of the matrix in the linear system is
on the order of the typical gap size between RNAPs during
transcription. The mean time for a backtracking polymerase to
incorporate the wrong nucleotide is $u(0,N)/v(0,N)$ when the initial
distance from the trailing polymerase is $N$. The corresponding mean
time for a backtracking polymerase to cleave the wrong nucleotide is
$\tilde{u}(0,N)/\tilde{v}(0,N)$ and the unconditioned mean escape time
is $u(0,N)+\tilde{u}(0,N)$. The above analyses provides an alternative
methods for computing mean exit times and have been verified against
the direct method presented in the main text.
\vfill
\eject
%
\end{document}